**Is the annual growth rate in Ireland's balance of trade time series nonlinear?**

Gerard Keogh

*Central Statistics Office, Dublin 6, Ireland*
Email: gerard.keogh@cso.ie

Abstract

We describe the Time Series Multivariate Adaptive Regressions Splines (TSMARS) method. This method is useful for identifying nonlinear structure in a time series. We use TSMARS to model the annual change in the balance of trade for Ireland from 1970 to 2007. We compare the TSMARS estimate with long memory ARFIMA estimates and long-term parsimonious linear models. We show that the change in the balance of trade is nonlinear and possesses weakly long range effects. Moreover, we compare the period prior to the introduction of the Intrastat system in 1993 with the period from 1993 onward. Here we show that in the earlier period the series had a substantial linear signal embedded in it suggesting that estimation efforts in the earlier period may have resulted in an over-smoothed series.

## 1   Introduction

The monthly balance of trade is the difference between the value of a country's merchandise exports and imports in current prices in a particular month. When seasonal variation and annual growth are taken into account, the market for traded goods that determines the balance of trade might reasonably be expected to comprise many competing agents, who have some flexibility to shift their demand (or supply) between time periods. Therefore it can be assumed that these agents behave rationally or, to be exact, the market will evolve according to the weak form of efficient market hypothesis (see Jasic and Wood 2006). In this environment the annual growth rate of the seasonally adjusted balance of trade is determined solely by overall demand and supply. In the short term, this implies that no agent can gain a significant advantage based on a study of autocorrelations or other information available from changes in values (i.e. prices). This of course implies the existence of a unit root in the level of the balance of trade series. Moreover, it suggests that annual changes in the balance of trade can have little or no predictability and so will evolve in a purely random manner. If this is the case the annual growth rate in balance of trade series can be said to be statistically in global balance as the overall flows into the economy balance the overall flows out.

In this article we examine this claim for the major world economies Australia, Canada, China, France, Germany, India, Italy, Japan, UK, US and Ireland– noting that the UK and US are Ireland's major trading partners. We ask, does the annual growth rate in the seasonally adjusted balance, denominated in US$, in each case evolve randomly and if not can we distinguish any particular features in the data. Two phenomena are of particular interest. One is the possibility of long memory or persistence in the data. Roughly speaking long memory means that events in the distance past effect current events. This is a weakly nonlinear behaviour prevalent in economic models for exchange rates (see Granger (1980), Hosking (1981), Diebold and Rudebush (1989) and Sowell (1992) to mention only a few).

A second possibility is that the underlying economy is more strongly nonlinear. Work in nonlinear modelling of economic data is less common than long memory work. Excluding the ARCH family, some notable exceptions within the parametric methods are the modelling of stock returns using state space Markov switching models (see Kim & Nelson 1999). Most of the effort in nonlinear modelling has centred on nonparametric tools (see Fan & Yao 2003). We adopt this approach and use TSMARS (Time Series Multivariate Adaptive Regressions Splines) due to Friedman (1991a) to model nonlinear effects in the growth rate in balance of trade series. We mention that TSMARS has also been used to model long range dependence (Lewis & Ray 1997) and exchange rates (De Goojer, Ray & Krager 1998) but other economic applications are unknown to us save Keogh (2006).

The plan of the remainder of the paper is as follows. In section 2 we provide some background on nonlinear models and self exciting threshold autoregressive models in particular. Section 3 describes TSMARS. In Section 4 we model the balance of trade series using long memory models and TSMARS. We contrast the resulting models for the balance of trade data. Section 5 concludes.

## 2 The growth rate in Balance of Trade

### 2.1 Data

The data we focus on are the official monthly balance of external trade in goods series denominated in US$ from 1970 to 2007 for the Irish economy –the series is obtained from the OECD database OECD.Stat. We denote the balance series by $y_t$ then given the weak form of efficient market hypothesis (Jasic and Wood 2006), a simple model for the balance series is to assume it follows a random walk with drift. More precisely we can say the annual growth rate series evolves according to

$$\frac{y_t - y_{t-12}}{y_{t-12}} = \varepsilon_t \qquad \varepsilon_t \sim N(0, \sigma^2) \tag{1}$$

Unfortunately this annual growth rate series is liable to nasty scaling problems that arise when $y_t$ is of moderate size but $y_{t-12}$ is relatively small. To avoid this problem we instead adopt the novel alternative and consider the symmetric annual growth rate in this series which we define as:

$$z_t = \frac{y_t - y_{t-12}}{|y_t| + |y_{t-12}|} \tag{2}$$

This series has the attractive feature that its maximum value is 1 while its minimum is –1. We are interested to see whether $z_t$ evolves randomly and if not then can we characterise the nature of the process defining $z_t$. A plot of $z_t$ for the Irish Balance of Trade series is given in Figure 1.

A striking feature of this plot is the smaller amount of variability in the series from about 1993 onward. This coincides with a change in the way estimates were compiled as a consequence of the introduction of the Single European Market and Intrastat system. In this system large traders make a detailed return of all imports and exports to the Irish Revenue Commissioners (Customs & Excise Office) while small traders are required only to record the total value of all goods imported or exported. It is clear from the plot that the Intrastat system has substantially smoothed the fluctuations in growth series $z_t$. Moreover, the presence of this discontinuity at 1993 is clear evidence that $z_t$ as it stands cannot be normally distributed. So, for European economies in the single market we propose to treat the series as two sub-series; the left portion prior to 1993 denoted by $z_{L,t}$ and the portion from 1993 onward denoted by $z_{R,t}$.

Figure 1: Symmetric annual growth rate in the Irish Balance of Trade series

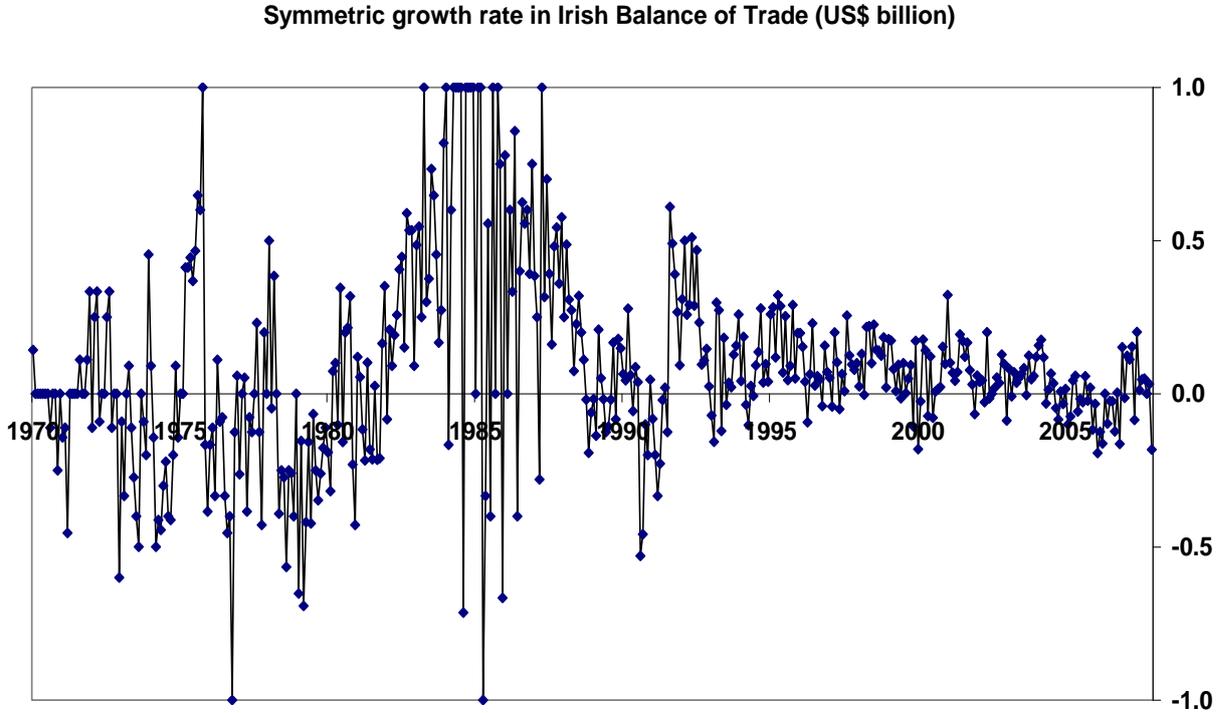

## 2.2 Autocorrelations

The autocorrelation function and two standard error (S.E.) limits for each of these series is plotted in Figure 2. It is clear that like many economic time series the autocorrelation functions decay slowly for both left and right sub-series indicating that the series have 'long memory' and so cannot evolve randomly.

## 3 Long Memory Modelling

If, after differencing, exponential decay is not observed in the autocorrelation function but rather it decays hyperbolically the time series is said to exhibit long memory. Time series models that describe this phenomenon are called fractionally differenced models. They generalise the usual concept of differencing by allowing the difference operator $(1-B)^d z_t$, ($B$ denotes the backward shift operator) applied to a time series $z_t$ to have non-integer values of differencing $0 < d < 1$. Using the Binomial Theorem we can readily expand the differencing operator as

$$(1-B)^d z_t = \left(1 - dB + \frac{1}{2!}d(d-1)B^2 - \frac{1}{3!}d(d-1)(d-2)B^3 + \ldots\right) z_t \qquad (3)$$

Figure 2: Autocorrelation Plot of Left and Right sub-series

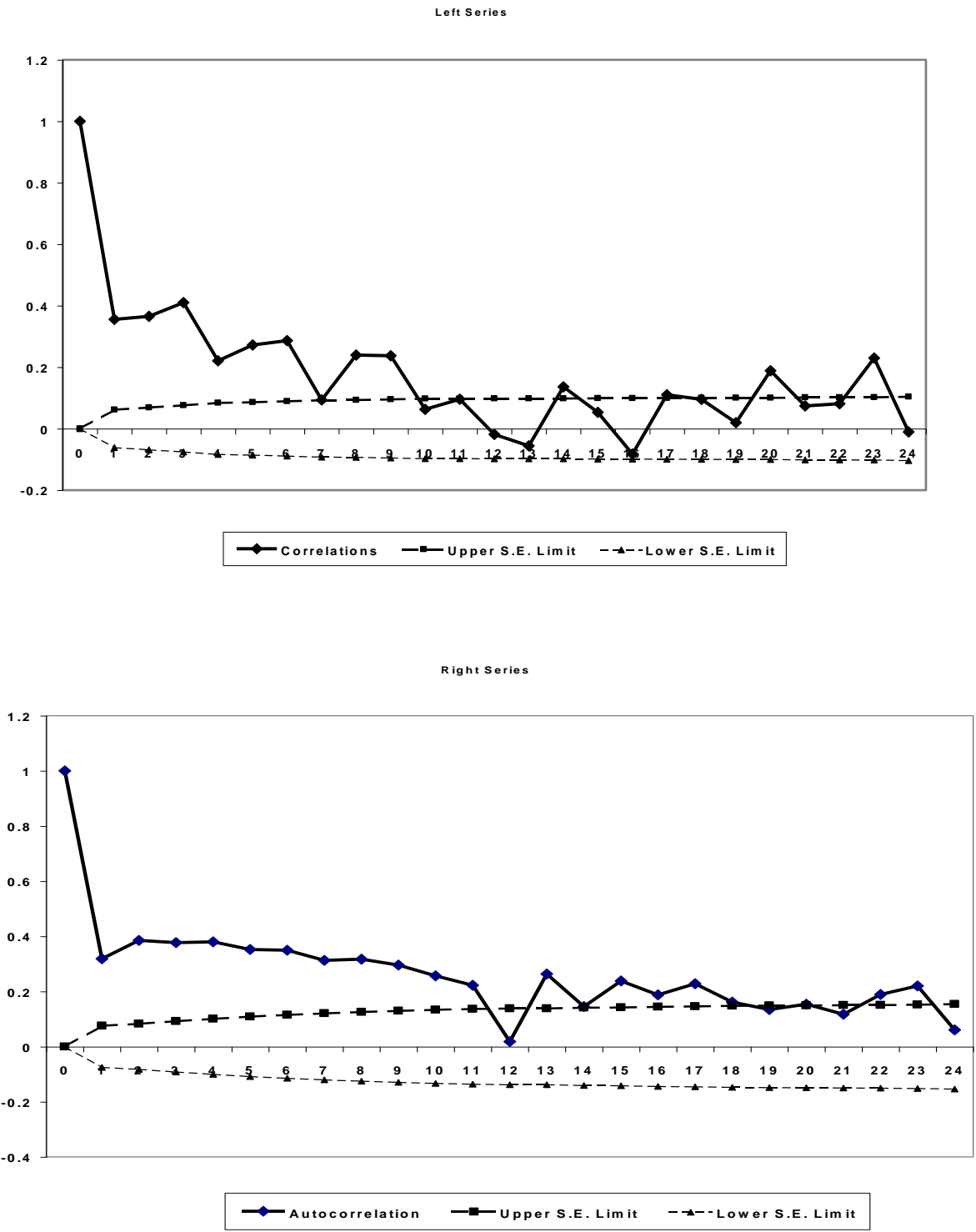

from which it is immediately clear that not only recent values but also values from the more distant past influence the current value. Such time series are said to have long memory. If $0 < d < 0.5$ the series is stationary while if $0.5 \leq d < 1$ the series is nonstationary. Furthermore, the class of 'Box-Jenkins' ARIMA$(p,d,q)$ models with autoregressive (AR) order $p$ and moving average (MA) order $q$ can also be generalised to include a fractional value for $d$, these models are called ARFIMA$(p,d,q)$ models (autoregressive fractionally integrated moving average models).

To the left and right sub-series respectively we fitted a sequence of ARFIMA$(p,d,q)$ models using the SAS function FARIMA with $0 \leq p,q \leq 3$ and computed the AIC criterion. For both sub-series we found that the best fitting model was the ARFIMA$(0,d,0)$ model with $d = 0.29$ for the left series and $d = 0.30$ for the right series. The autocorrelation functions of the resulting filtered series are plotted in Figure 3. Both plots show that the slow decay has been eliminated by the fractional differencing. However, the left series plot displays a number of autocorrelations lying outside the two standard error limits while the right series plot shows a significant autocorrelation at the seasonal lag (period twelve). This suggests that the data do possess long memory properties but that fractionally differenced models are insufficient to explain all the structure in the data

Figure 3: Autocorrelation Plot of Fractionally differenced Left and Right subseries

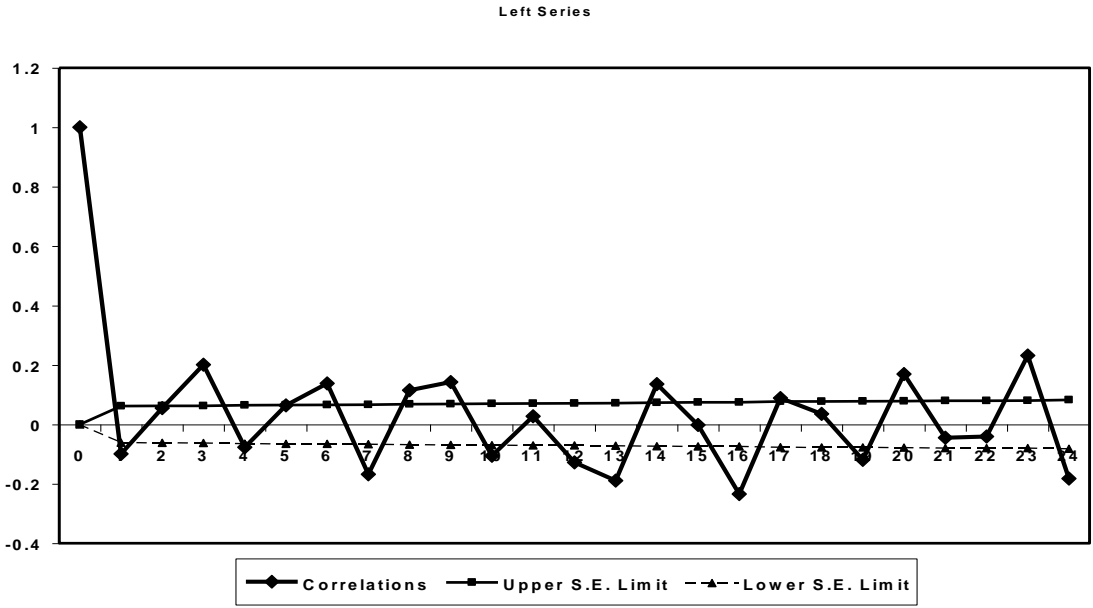

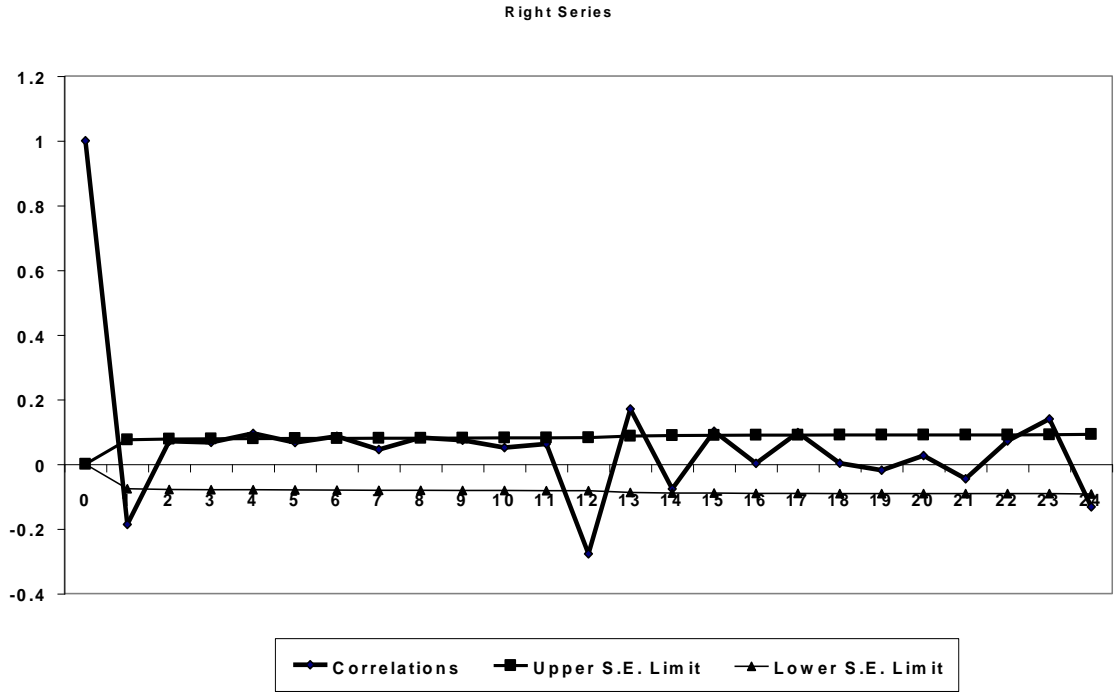

**Right Series**

## 4  Nonlinear modelling with TSMARS

### The TSMARS model

Multivariate Adaptive Regressions Splines (MARS) is a flexible nonparametric smoothing procedure due to Friedman (1991a) with extensions described in Friedman (1991b, c). MARS seeks to approximate the underlying functional relationship between a single response variable $y$ and a vector $p$ of predictor variables $\mathbf{x} = (x_1, \ldots x_p)$ through the nonlinear regression equation

$$y = f(x_1, \ldots x_p) + \varepsilon \tag{4}$$

over some domain $D \subset \mathbf{R}^p$ that contains n samples of data, namely $\{y_i, \mathbf{x}_i\}_{i=1}^n$ and having i.i.d. error $\varepsilon$. Specifically, when $y_i = y_t$ is the $t^{\text{th}}$ value in the time series $\{y_t\}$ and we let the predictors be the lagged values $y_{t-1}, y_{t-2}, \ldots, y_{t-p}$ we get TSMARS. In TSMARS the regression function $f(\bullet)$ is modelled according to

$$\hat{y}_t = \hat{f}_t = \beta_0 + \sum_{m=1}^{M} \beta_m \prod_{k=1}^{K_m} \left[ s_{k,m} \left( y_{v(k,m)} - y^*_{\xi(k,m)} \right) \right]_+ \tag{5}$$

This is a linear combination of M+1 basis functions $\prod_{k=1}^{K_m}\left[s_{k,m}\left(y_{v(k,m)} - y^*_{\xi(k,m)}\right)\right]_+$, with each of these a tensor product of $K_m$ linear spline functions. The quantity $K_m$ is the interaction degree of the spline functions and is normally limited to 3. Each spline function $s_{k,m}\left(y_{v(k,m)} - y^*_{\xi(k,m)}\right)$ comprises the sign denoted by $s_{k,m}$, a single lagged predictor variable labelled $y_{v(k,m)}$ and knot or threshold point $y^*_{\xi(k,m)}$ that splits the domain of $y_{v(k,m)}$ into two sub-regions according to the sign of the subscript. Thus, for example, a "+" subscript indicates the positive part of the argument, i.e. $[z]_+ = \begin{cases} z, & \text{if } z > 0 \\ 0, & \text{otherwise} \end{cases}$. MARS is a so-called greedy algorithm that approximates $f(\bullet)$ by starting with a basis function set that consists of the single constant basis function $b_0 = 1$. It then exhaustively searches over the existing set of basis functions, followed by the set of predictor variables, followed by all values (i.e. knots) of that predictor variable. This generates a new basis function that is added to the existing set on the basis that it is a) orthogonal to the existing set; and b) has minimum Generalised Cross Validation (GCV) lack-of-fit score among all basis functions in the available search space. We mention the GCV is basically the residual sum of squares (RSS) associated with the regression of the set of basis functions on the response $y_t$, that penalised for model complexity with a smoothing parameter $d$ which is set by the user. The forward search is followed by a backward "pruning" of the basis functions model. This procedure generates a "good" set of basis functions for approximating the underlying function.

The nonlinear time series model defined by (5) is known as the Adaptive Spline Threshold Autoregressive (ASTAR) model of Lewis & Stevens (1991) – following Lewis & Ray (1997) we synonymously call this a TSMARS model. This model, as pointed out by Lewis & Stevens (1991), admits the $l$-regime SETAR$(l, p_1, p_2, \ldots p_l)$ model of Tong (1990) as a special case; this being obtained by simply restricting the interaction degree of the spline basis functions to $K_m = 1$. For more general nonlinear models the advantage of TSMARS is that it admits continuous models with possibly more than 1 threshold, see Lewis & Stevens (1991).

**ANOVA Decomposition**

As noted in Friedman (1991a), the constructive representation given by equation (5) provides little insight into the nature of the approximation. However, by simply rearranging terms, (5) can be cast into form that reveals a lot about the predictive relationship between the response and lagged co-variates. This allows us to partition the variance of the TSMARS model according to basis function type. A straightforward computation then gives the percentage of the Total Sum of Squares (TSS) attributable to nonlinear basis functions (see Keogh 2006).

The basic idea is to collect together all linear basis functions, second order tensor product basis functions, all third order terms and so on. An example will best illustrate the idea. Take the following TSMARS approximation based on the lagged co-variates denoted for convenience by $x_1 = y_{t-1}, x_2 = y_{t-2}$, and $x_3 = y_{t-3}$

$$\hat{f} = 0.05 + 0.1 y_{t-1} - 0.5 y_{t-3} - 0.18 (y_{t-1} - 0.35)_+ + 0.3 y_{t-1}(y_{t-2} - 0.5)_- + 0.2 y_{t-2} y_{t-3} - 0.1 y_{t-1}(y_{t-2} - 0.5)_-(y_{t-3} - 0.3)_+$$

This TSMARS model can be rewritten as an additive combination of the form

$$\hat{f} = \begin{matrix} f_0^* + \\ f_1^*(y_i) + \\ f_2^*(y_i, y_j) + \\ f_3^*(y_i, y_j, y_k) \end{matrix} \begin{matrix} = 0.05 + \\ = 0.1 y_{t-1} - 0.5 y_{t-3} - 0.18 (y_{t-1} - 0.35)_+ + \\ = 0.3 y_{t-1}(y_{t-2} - 0.5)_- + 0.2 y_{t-2} y_{t-3} - \\ = 0.1 y_{t-1}(y_{t-2} - 0.5)_-(y_{t-3} - 0.3)_+ \end{matrix}$$

This is a sum over all basis functions that involve a single variable only, two variables only and involving three variables only. Owing to the similarity of this expression with the decomposition of the analysis of variance for contingency tables, Friedman (1991a) refers to it as the ANOVA decomposition of the model (5). Moreover, due to the fact that the basis functions are orthogonal we can formally partition the overall variance (i.e. total sum of squares) see Keogh (2006) as follows:

$$\begin{aligned} V(y) &= V(\hat{f}) + V(\text{residual}) \\ &= V(f_0^*) + V(f_1^*(y_{t-i})) + V(f_{1,S}^*(y_{t-i})) + V(f_2^*(y_{t-i}, y_{t-j})) + V(f_3^*(y_{t-i}, y_{t-j}, y_{t-k})) + \ldots + V(\text{residual}) \\ &= V(\text{constant}) + V(\text{linear}) + V(\text{linear spline}) + V(\text{nonlinear}) + V(\text{residual}) \end{aligned} \quad (6)$$

where $V(\bullet)$ is the variance. This partition of the overall variance is crucial, as it provides a measure of the extent of nonlinearity in a given set of data. In addition it allows us to express the contribution of nonlinear basis functions to the overall variance as

$$\% \text{Nonlinearity} = \frac{V(\text{nonlinear})}{V(y)} \times 100\%$$

### 4.1 TSMARS modelling

From equation (3) it is clear that the fractionally differenced model can account for long range effects by incorporating many lags into the AR polynomial. This model is a weakly nonlinear generalisation of the linear ARIMA model and therefore, may not fully capture the dynamics of the symmetric annual growth rate in the balance of trade series $z_t$. We propose to tackle this issue by analysing $z_t$ using TSMARS with many lagged predictors included to account for long range effects.

Initially we identify the best model by fitting a sequence of TSMARS models to the left $z_{L,t}$ and right $z_{R,t}$ sub-series respectively. Our procedure involves producing a set of fitted models using $p = $ 6, 12, 24, 36, 48 and 60 lagged predictors respectively for each sub-series. We also allow the maximum interaction degree $\kappa_m$ of the basis function splines to equal one (linear segments) or two (quadratic segments) respectively. For each sub-series this equates to twelve runs of the TSMARS program and each run results in a fitted model. In each case the smoothing parameter $d = 3$ associated with the GCV and the maximum number of allowable basis functions is set equal to number of predictors $p$.

Table 1: TSMARS model fitting results for left sub-series

| | | | | No of model parameters | | ANOVA Decomposition % of Total Sum of Squares explained by each type of basis function | | | |
|---|---|---|---|---|---|---|---|---|---|
| $p$ | $\kappa_m$ | RSS | GCV | Linear | Non-linear | Constant | Linear | Non-linear | Residual |
| 6 | 1 | 35.4 | 37.7 | 2 | 0 | 4.9 | 22.1 | 0.0 | 73.0 |
| | 2 | 36.7 | 38.3 | 2 | 0 | 2.3 | 22.1 | 0.0 | 75.6 |
| 12 | 1 | 30.4 | 34.2 | 1 | 2 | 2.2 | 18.8 | 16.3 | 62.7 |
| | 2 | 29.2 | 35.3 | 1 | 6 | 2.2 | 16.8 | 20.8 | 60.2 |
| **24** | **1** | **24.0** | **31.5** | **5** | **7** | **2.3** | **24.9** | **23.4** | **49.4** |
| | 2 | 23.2 | 36.2 | 2 | 12 | 2.2 | 20.1 | 29.9 | 47.8 |
| 36 | 1 | 22.1 | 31.6 | 11 | 7 | 3.1 | 28.0 | 23.4 | 45.5 |
| | 2 | 23.8 | 49.1 | 2 | 9 | 2.9 | 18.5 | 29.5 | 49.1 |
| 48 | 1 | 19.7 | 33.0 | 12 | 9 | 2.2 | 29.6 | 27.7 | 40.5 |
| | 2 | 21.5 | 84.9 | 2 | 13 | 4.5 | 18.2 | 33.1 | 44.2 |
| 60 | 1 | 19.7 | 36.1 | 18 | 9 | 0.0 | 36.1 | 28.0 | 35.9 |
| | 2 | 21.2 | 80.6 | 5 | 11 | 3.4 | 29.6 | 23.4 | 43.6 |

The results of this procedure are displayed in Tables 1 and 2 for the left and right sub-series respectively. In the tables the RSS and GCV for each model are given. In columns five and six we provide a breakdown of the number of model parameters attributable to linear and nonlinear basis function sets (note: there is always only one constant basis function). The remaining columns give the Total Sum of Squares (TSS) explained by the respective basis function sets and the percentage remaining in the residual.

The results for the left sub-series are clear. As the number of lagged predictors allowed into the model increases the RSS falls quite fast initially and then levels off to about twenty. This fall is occurs irrespective of whether the maximum interaction degree $\kappa_m$ of the basis function splines is one or two. By contrast the GCV initially falls and then begins to rise again as the penalty for more complex models increases. Both the behaviour of the RSS and GCV indicate that TSMARS has performed smoothly when modelling these data. The minimum GCV occurs at a lag $(p)$ of twenty-four and maximum interaction degree of $(\kappa_m)$ of one.

There are five linear and seven (non) linear spline segments in the model showing the chosen model is of medium complexity from among the twelve fitted. These elements account for 24.9 and 23.4 percent of the TSS respectively. We note in this case the nonlinear basis functions are in fact linear splines where the knot is within the interval of the data so there is a left and right basis function associated with that knot. Meanwhile the purely linear spline has its knot at the end point of the data, so in this case there is just a single (left) linear basis function predictor. With nonlinear basis functions accounting for 23.4% of the TSS and nonlinear basis functions also contributing to all models using twelve or more lagged predictors we conjecture the left sub-series is nonlinear and the nonlinearity is approximated well with a linear combination of linear spline segment functions.

The results for the right sub-series are not as clear-cut as those for the left series. Once again as the number of lagged predictors in the model increases the RSS falls quite fast initially but then begins to level off to settle in the range of about 0.8 to 1.03. Moreover, this behaviour is somewhat more erratic when the maximum interaction degree $\kappa_m$ of the basis function splines is two. The movement of the GCV when maximum interaction is one starts at about 1.8 and rises with more lagged predictors, it then starts to fall when forty-eight or more lagged predictors are included in the model. With sixty lagged predictors and a maximum interaction of one we get the lowest GCV model. In contrast when the maximum interaction is equal to two TSMARS - produces models with larger GCV values. Indeed when the lag is thirty-six or more TSMARS tends to over penalise the more complex models. In this case the behaviour of the RSS and GCV indicate that the response of TSMARS is uneven when modelling these data. That said, it is clear that the minimum GCV occurs at a lag $(p)$ of sixty and maximum interaction degree of $(\kappa_m)$ of one. In this case there are three purely linear and ten linear (split) spline segments in the model. Clearly the chosen model is among the most complex of the twelve models fitted. The linear and linear spline segments elements account for five and 52.7 percent of the TSS respectively. Furthermore, since nonlinear basis functions make a substantial contribution to all models fitted we conjecture the right sub-series is nonlinear and once again the nonlinearity is suitably approximated with a linear combination of linear spline segment functions. Further analysis proceeds using the five linear and seven linear spline segments model for the left sub-series and the three linear and ten linear spline segments model for the right sub-series.

Table 2: TSMARS Model fitting results for right subseries

|  |  |  |  | No of model parameters | | ANOVA Decomposition % of Total Sum of Squares explained by each type of basis function | | | |
| --- | --- | --- | --- | --- | --- | --- | --- | --- | --- |
| $p$ | $\kappa_m$ | RSS | GCV | Linear | Non-linear | Constant | Linear | Non-linear | Residual |
| 6 | 1 | 1.59 | 1.77 | 2 | 2 | 15.5 | 6.6 | 24.8 | 53.1 |
|  | 2 | 1.65 | 1.89 | 1 | 3 | 19.9 | 4.5 | 20.5 | 55.1 |
| 12 | 1 | 1.42 | 1.70 | 3 | 4 | 17.6 | 7.7 | 27.5 | 47.2 |
|  | 2 | 1.36 | 1.86 | 1 | 8 | 17.8 | 2.7 | 34.2 | 45.3 |
| 24 | 1 | 1.11 | 1.83 | 7 | 3 | 17.0 | 20.1 | 25.9 | 37.0 |
|  | 2 | 0.85 | 1.39 | 1 | 3 | 29.8 | 13.4 | 28.4 | 28.4 |
| 36 | 1 | 0.91 | 2.41 | 6 | 9 | 13.4 | 14.2 | 42.0 | 30.4 |
|  | 2 | 1.03 | 3.21 | 3 | 8 | 13.3 | 11.5 | 41.0 | 34.2 |
| 48 | 1 | 0.87 | 1.57 | 3 | 9 | 13.0 | 12.6 | 45.5 | 28.9 |
|  | 2 | 1.03 | 3.09 | 5 | 6 | 9.5 | 17.8 | 38.4 | 34.3 |
| **60** | **1** | **0.80** | **1.16** | **3** | **10** | **11.0** | **5.0** | **57.2** | **26.8** |
|  | 2 | 0.84 | 2.85 | 8 | 4 | 0.4 | 16.2 | 55.3 | 28.1 |

## 4.2 The left sub-series model

The model TSMARS gave for the left sub-series is based on a lag length of twenty-four and the maximum interaction degree $\kappa_m$ of the basis function splines is one. The full model is:

$$\hat{z}_t = -0.673 + 0.262\, z_{t-2} + 0.134\, z_{t-3} + 0.090\, z_{t-8} - 0.176\, z_{t-16} + 0.188\, z_{t-20} -$$
$$0.437\, (z_{t-1} - 0.189)_- + 0.361\, (z_{t-9} + 0.135)_+ + 0.388\, (z_{t-12} - 0.438)_- + 0.318\, (z_{t-14} - 0.003)_+ +$$
$$0.167\, (z_{t-18} + 0.446)_+ - 0.450\, (z_{t-19} + 0.445)_+ - 0.437\, (z_{t-24} - 0.350)_+$$
(7)

From the ANOVA decomposition we know that nonlinearity accounts for 23.4% of the TSS, this is clearly substantial. Interestingly, by way of comparison we could find no simple linear AR model that fitted these data well. However certain linear seasonal AR models were found to fit the data quite well, an example is

$$\hat{z}_t = 0.227\, z_{t-1} + 0.236\, z_{t-2} + 0.231\, z_{t-3} + 0.161\, z_{t-9} + 0.269\, z_{t-12}$$
$$- 0.200\, z_{t-14} + 0.145\, z_{t-16} + 0.238\, z_{t-20} - 0.146\, z_{t-24} - 0.136\, z_{t-48} + \varepsilon_t$$
(8)

In Figure 4 we compare the autocorrelation plot of the TSMARS model with an AR (seasonal) model. The comparison is striking. Both models give a sequence of autocorrelations that decay rapidly. Up to lag twenty-four TSMARS does marginally better than the AR model. Both methods have some difficulty coping with the correlations at the seasonal lags of twenty-four and thirty-six, though TSMARS does marginally better. The complexity in terms of the number of parameters and the lagged predictors involved is strikingly similar but it is significant that TSMARS does not include lags beyond twenty-four. The inclusion of the predictor at lag forty-eight in the linear model is important but its exclusion from the TSMARS model does not seem to affect the autocorrelation. This shows that medium range effects are important in both models. Looking at the model equation (7), medium range dependence is provided for by including lags sixteen and twenty linearly in the model. In addition nonlinear medium range effects are provided for by the linear spline functions at lags eighteen and nineteen. These two basis functions account for only three percent of the variance but are important as they make a contribution to the fit only when the $z_{t-18,19} + 0.446 > 0$. This effect becomes more pronounced at $z_{t-9} = 0.135$. Nonlinear seasonal effects are included at lags twelve and twenty-four, these account for a substantial seven percent of the variance but only influence the fit when $z_{t-12,24} > 0.35$, causing a nonlinear downward movement in the seasonal response at these lags.

Figure 4: Autocorrelation plots for models of the Left sub-series

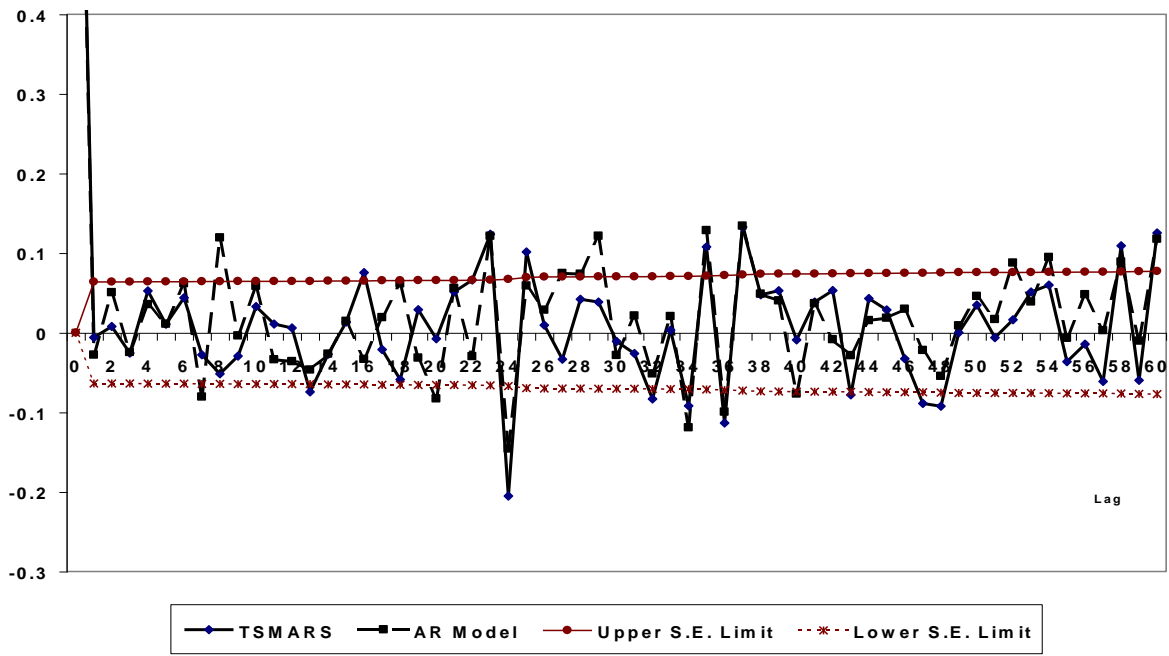

A little more insight is provided when we transform back via equation (2) and consider what happens in the original balance of trade variable $y_t$. For example, in the period prior to 1983 the balance is almost continuously negative. Putting $x_t = |y_t|$ the spline term $z_{t-18} + 0.446 > 0$ which produces the medium range curvature, is transformed to $x_{t-18} < 2.61\, x_{t-30}$; $z_{t-19}$ is transformed similarly. This says that movement only occurs in the process when the absolute value of the series is over 2.6 times lower than the previous year's value. This happened in fourteen of the months up to 1983 and typically these occurred in the third quarter of the year. The worst year for the balance of trade was 1979 where the fall more than doubled relative to 1978. Our model shows this sharp fall affected the balance of trade up to two-and-half years after the initial shock. Meanwhile, a portion of this downward pressure is relieved by seasonal movements in the trade when the terms $z_{t-12,24} > 0.35$ is transformed to influence the outcome as an upward movement.

We also conducted a normality test on the residuals and this gave a p-value of 0.06 while Q-Q plots showed no obvious deviation from normality in the residuals. Furthermore, an autocorrelation plot of the squared residuals also showed no evidence of autocorrelation indicating that data are not hetroscedastic. These facts strongly point to the conclusion that these data are not evolving in a purely random way but are generated by a weakly nonlinear medium range dependent process. Of course we mention that this process may have been an artificial by-product of procedures used to compile the external trade data in the period up to 1993.

### 4.3 The right sub-series model

The model TSMARS gives for the right sub-series is based on a lag length of sixty and the maximum interaction degree $\kappa_m$ of the basis function splines is one. The full model is:

$$\begin{aligned}\hat{z}_t = &-0.051 - 0.135\, z_{t-16} + 0.221\, z_{t-36} - 0.197\, z_{t-49} - \\ & 0.296\, (z_{t-2} - 0.149)_+ + 0.3441\, (z_{t-3} + 0.017)_+ + 0.448\, (z_{t-7} - 0.021)_+ + \\ & 0.478\, (z_{t-12} - 0.103)_- - 0.316\, (z_{t-17} - 0.093)_- - 0.124\, (z_{t-35} - 0.236)_- - \\ & 0.857\, (z_{t-35} - 0.236)_+ - 0.903\, (z_{t-40} - 0.236)_+ - 0.441\, (z_{t-47} - 0.113)_- - 0.109\, (z_{t-50} + 0.051)_+\end{aligned} \quad (9)$$

The curvilinear nature of the data is clear from the fact that 57.2% of the variance is explained by nonlinear basis functions in the ANOVA decomposition in Table 2. Moreover, short range dependence is particularly important as the two basis functions at lags two and three account for 28.5 and 9.7% of the variance respectively. Clearly these elements are producing most of the curvature associated with this model. Transforming the lag two term $z_{t-2} - 0.149 > 0$ back via equation (2) we can see the effect of this term on the original balance of trade variable $y_t$. Throughout the period from 1993 onward the balance is continuously positive. Putting $x_t = |y_t|$ this spline term is transformed to $x_{t-2} > 1.35\, x_{t-14}$. Therefore the right hand trade process responds in a nonlinear manner when the value of the series is over 1.35 times its previous year's value. This happened in on thirty-seven occasions after 1993. Among these the year 1998 stands out with nine of the months showing a thirty five percent growth on 1997. This coincides with a period of unparalleled growth in Irish trade where year-on-year constant price changes of well over twenty percent. Therefore, our model shows that, trade in 1998 was largely influenced by a one-off short–term ramp that marked the start of new phase in Irish economic evolution.

Medium range dependence in equation (9) is provided for linearly and nonlinearly at lags sixteen and seventeen in the model. Longer range effects are provided for at lags 35, 40, 49 and 50. In all these account for only six percent of the variance; 1.4% of which is linear at lag forty-nine and the remainder is comprised of linear splines with a knot (threshold) at $z_t = 0.236$. This relatively small amount of long range behaviour is important to the quality of the model as without these terms long range effects are not eliminated in the autocorrelation function. Once again we applied a normality test on the residuals and this gave a p-value of 0.94 while Q-Q plots showed no obvious deviation from normality in the residuals.

By way of contrast we searched for and found a linear seasonal model that also fits the data well is

$$\hat{z}_t = 0.386\, z_{t-2} + 0.33\, z_{t-6} - 0.220\, z_{t-12} - 0.220\, z_{t-24} + \varepsilon_t \tag{10}$$

A comparison of the autocorrelations from the residuals of the TSMARS and linear seasonal models are shown in Table 3. For TSMARS residuals there is no evidence of autocorrelation. In contrast the linear model shows significant autocorrelation in the residuals. Autocorrelation checks were also carried out on the squares of the residuals. Once again the tests provided no evidence of autocorrelation for the TSMARS model but the linear seasonal model failed the test at lags one and two. This departure from randomness would appear to be picked up by the lag two and three spline segments in model (9). Therefore we conclude the right sub-series $z_{t,R}$ is not evolving in a purely random manner from 1993 onward but possesses short-term nonlinear with weakly nonlinear long range effects. Thus, since the introduction of the Intrastat system, the symmetric annual growth in the balance of trade has been characterised by short-term nonlinear movements that were not evident prior to the 1993.

Table 3: Autocorrelation check for TSMARS Model fit of right sub-series

| Lag | $\chi^2$ statistic | Degrees of freedom | $\chi^2$ probability | Autocorrelations | | | | | |
|---|---|---|---|---|---|---|---|---|---|
| TSMARS Residual | | | | | | | | | |
| 6 | 8.1 | 6 | 0.23 | -0.16 | -0.10 | -0.14 | 0.11 | -0.01 | 0.02 |
| 12 | 12.87 | 12 | 0.38 | -0.03 | 0.02 | 0.06 | -0.09 | 0.04 | -0.15 |
| 18 | 16.03 | 18 | 0.59 | 0.06 | 0.05 | 0.11 | -0.07 | -0.01 | -0.02 |
| 24 | 21.88 | 24 | 0.59 | 0.04 | 0.02 | -0.07 | 0.03 | 0.05 | -0.17 |
| Linear Seasonal AR Model Residuals | | | | | | | | | |
| 6 | 23.16 | 6 | <0.01 | 0.02 | -0.212 | 0.116 | 0.204 | 0.129 | -0.102 |
| 12 | 35.15 | 12 | <0.01 | 0.045 | 0.08 | 0.076 | 0.083 | -0.051 | -0.199 |
| 18 | 44.55 | 18 | <0.01 | 0.109 | 0.028 | 0.042 | 0.01 | 0.081 | 0.163 |
| 24 | 61.72 | 24 | <0.01 | -0.033 | 0.018 | -0.041 | 0.163 | 0.143 | -0.18 |

## 5 Conclusions

We have set out the Time Series Multivariate Adaptive Regressions Splines (TSMARS) method. This method is useful for identifying nonlinear structure in a time series. In particular it is useful in identifying SETAR and ASTAR models. We have used TSMARS to model the annual change in the balance of trade for Ireland in the year 1970 to 2007. Using the GCV model selection criterion and the ANOVA decomposition available in TSMARS we identified a 'best' fitting model. We compared the resulting TSMARS estimate with long memory ARFIMA estimates and long-term parsimonious linear models and showed that TSMARS found both short-term nonlinear effects and weaker long range effects. In this way the TSMARS estimate has reproduced and improved upon the estimates of linear and medium range effects that were evident from the best linear and fractionally diffferenced models.

From an economic perspective it is clear from the results that the change in the balance of trade is not evolving randomly but is in fact nonlinear. In particular we have shown that the introduction of the Intrastat system for compiling trade figures coincided with a change in the evolution of the series. Prior to 1993 the series possessed a strong linear signal that accounted for about one quarter of the overall variance coupled with a weakly linear medium memory structure. However, from 1993 the introduction of the Intrastat system has coincided with a shift to a nonlinear ramp-type short-term evolution of the balance of trade series coupled with weakly linear effects. The explanation for this change is a matter for further study. Nonetheless, one conjecture worth noting is that the Intrastat system eliminated issues pertinent to the timing of transactions. In the years prior to 1993 estimation associated with these transactions may have smoothed the data and thereby artificially strengthened the linear signal in the data.